\documentclass[]{interact}

\usepackage{epstopdf}
\usepackage[caption=false]{subfig}

\usepackage[numbers,sort&compress,merge]{natbib}
\usepackage{hyperref}
\usepackage{threeparttable}
\bibpunct[, ]{[}{]}{,}{n}{,}{,}
\theoremstyle{plain}

\theoremstyle{definition}

\theoremstyle{remark}

\usepackage{indentfirst}
\usepackage[version=3]{mhchem} 
\newcommand{\lambdabar}{{\mkern0.75mu\mathchar '26\mkern -9.75mu\lambda}}
\usepackage{xcolor}

\begin{document}

\articletype{ARTICLE TEMPLATE}

\title{Towards highly accurate calculations of parity violation in chiral molecules: relativistic coupled-cluster theory including QED-effects}

\author{
\name{
Ayaki Sunaga\textsuperscript{a} 
and 
Trond Saue\textsuperscript{b}\thanks{CONTACT T. Saue. Email: trond.saue@irsamc.ups-tlse.fr}}
\affil{
\textsuperscript{a}Institute for Integrated Radiation and Nuclear Science, Kyoto University, 2 Asashiro-Nishi, Kumatori-cho,
Sennan-gun, Osaka 590-0494, Japan\\
\textsuperscript{b}Laboratoire de Chimie et Physique Quantique, UMR 5626 CNRS--Universit{\'e} Toulouse III-Paul Sabatier, 118 Route de Narbonne, F-31062 Toulouse, France}
}

\maketitle

\begin{abstract}
Parity-violating energies $E_{\rm{PV}}$ of the \ce{H2X2} (X = O, S, Se, Te, Po) molecules are reported, calculated as analytical expectation values at the relativistic coupled-cluster singles-and-doubles (CCSD) level using property-optimized basis sets. Radiative corrections to the $E_{\rm{PV}}$ was investigated using effective QED-potentials and found to reach a maximal value of 2.38\% for \ce{H2Po2}. However, this result depends on the choice of effective self-energy potential and may indicate limitations to their domain of validity.
\end{abstract}


\section{Introduction}

In early 2012 the journal \textit{Nature} had a News Feature article on ``tough science'' listing five experiments ``as hard as finding the Higgs'' \cite{Jones_Nature2012}. One experiment in the list attempts to provide the first observation of parity violation (PV), associated with the weak force, in molecular systems using high-resolution laser spectroscopy\cite{Crassous_OBC2005}.
It is an example of a spectroscopic test of fundamental physics in the low-energy regime, using atoms and molecules, in vivid contrast to the Large Hadron Collider, 27 km long and operating at several TeV \cite{Erler_PPNP2005,Safronova_RevModPhys.90.025008}. 

It was Lee and Yang who in 1956 pointed out that there was insufficient experimental evidence for the conservation of parity in processes involving the
weak interaction and proposed experiments to test possible parity violation\cite{Lee_Yang_PhysRev.104.254}, soon to be confirmed by experiments\cite{wu1957experimental,Postma_Physica1957,Garwin_PhysRev.105.1415,Friedman_PhysRev.105.1681.2}. The processes that were studied are mediated by \textit{charged} vector bosons, in contrast to the neutral photon of electromagnetic interactions, and are therefore incompatible with stable atoms and molecules.
The situation changed with the development of electroweak theory\cite{Glashow_RevModPhys.52.539,Weinberg_RevModPhys.52.515,Salam_RevModPhys.52.525} and in particular the prediction of the existence of a neutral partner, $Z^0$, of the charged $W^\pm$ bosons. Early studies about the feasibility of observing parity violation in stable atoms were rather pessimistic\cite{Zeldovich_JETP1959,Michel_PR1965}, but limited attention to the hydrogen atom. M. A. and C. Bouchiat noted that PV has a strong scaling with nuclear charge and therefore suggested to look for such effects in heavy atoms\cite{Bouchiat_PLB1974,Bouchiat_JPF1974} and, indeed, the first observation of atomic PV was in the form of optical activity in a vapor of bismuth atoms\cite{Barkov_JETPlett1978,Barkov_PLB1979}. Since then PV has been observed in several heavy atoms:  thallium\cite{Conti_PRL1979,Bucksbaum_PRL1981,Drell_PRL1984,Vetter_PRL1995,Edwards_PRL1995},  cesium\cite{Bouchiat_PLB1982,Bouchiat_JPF1986,wood1997measurement,Bennett_PRL1999,Guena_PRL2003,Guena_PRA2005},  lead\cite{Emmons_PRL1983,Meekhof_PRL1993,Phipp_JPB1996}, bismuth\cite{Macpherson_PRL1991} and ytterbium\cite{Tsigutkin_PRL2009,Antypas_NaturePhysics2019}. For a recent review on atomic parity violation, see Ref.~\citenum{Roberts_ARNPS2015}.

For chiral molecules parity violation induces a finite, albeit minute energy difference between enantiomers\cite{letokhov1975difference,zel1977energy,Rein_PLA1979}, making them formally diastereomers, an intriguing fact in view of the observed chiral preferences in biological systems\cite{Bonner_TSC1988,Quack_Ang2002,Tsarev2009PPN}, but yet to be observed. The experiment under construction in Paris will search for the signature of parity violation in the form of a frequency shift in the vibrational spectra of left- and right-handed molecules. It aims for a measurement precision below 0.1 Hz and is based on the method of the Doppler-free two-photon Ramsey fringes in a supersonic molecular beam\cite{Crassous_OBC2005,Darquie:2010:chirality}. Recent progress is reported in Ref.~\citenum{Cournol_QE2019}, whereas information about other experiments and  molecular (as well as atomic) PV in general is found in an excellent review by Berger and Stohner\cite{Berger2019WCM}.

The role of theory at this point is to guide experiment. We have developed a computational protocol for the rapid screening of candidate molecules, based on 2-component relativistic density functional theory (DFT)\cite{DeMontigny_PCCP2010}. However, calculations of higher accuracy are needed, in the short run to calibrate the current computational protocol for molecular PV calculations, in the longer run to confront a possibly successful experiment. In this context coupled cluster theory stands out as the method of choice\cite{Bartlett_Stanton_RCC1994,Crawford_Schaefer_RCC2000,Bartlett_RevModPhys.79.291,Liu_Cheng_WIRE2021}. In 2000 Thyssen \textit{et al.} reported 4-component wavefunction-based correlated calculations of the parity-violating energy of the \ce{H2O2} and \ce{H2S2} molecules using a finite-field
approach at the level of 2nd-order M{\o}ller--Plesset perturbation theory (MP2), coupled-cluster singles-and-doubles (CCSD), CCSD with approximate triples correction (CCSD(T)) as well as configuration-interaction singles-and doubles (CISD)\cite{Thyssen_PRL2000}. The \ce{H2X2}-series of molecules are not good candidates for a PV experiment due to the almost free rotation around the X--X bond, but have been much used for benchmarking and analysis purposes due to their simple structures (see for instance Refs.\cite{mason1984parity,Bakasov_JCP_1998,lazzeretti1997calculation,lazzeretti1997calculation,laerdahl1999fully,Hennum_CPL2002,Berger_vanWullen_JCP2005,Berger_JCP2008,bast2011analysis,Horny_MP2015}). In 2005 van Stralen \textit{et al.} reported the first implementation of analytic one-electron
properties at the 4-component relativistic MP2 level and reported PV energies for \ce{H2X2}, (X=O,S,Se,Te) \cite{van2005first}. More recently, in 2016, these analytic calculations were extended to the CCSD level by Shee \textit{et al.} \cite{shee2016analytic}.

In the present work we repeat the calculations of Shee \textit{et al.} \cite{shee2016analytic}, but with a number of improvements: i) The coupled cluster calculations have been carried out using a new coupled cluster code, ExaCorr, implemented for massively parallel GPU-accelerated computing architectures\cite{pototschnig2021implementation}. ii) We have extended the Dyall-type all-electron basis sets with tight exponents optimized for PV calculations. iii) Using \ce{H2Se2}, as detailed in the Supplemental Material, we have investigated the effect of truncation of the occupied and virtual spaces  on the PV energy. iv) We have extended the calculations to include \ce{H2Po2}. v) Most importantly, we have to our knowledge carried out the first study of the effect of quantum electrodynamics (QED) on molecular PV energies.

We are here interested in the electron self-energy (SE) and the vacuum polarization (VP), both at the origin of the Lamb shift\cite{lamb1947fine}. QED, as expressed through the scattering matrix formalism, can describe atoms and molecules to amazing precision, but is in practice limited to few-electron systems, in particular due to the slow convergence of electron correlation within this formalism\cite{Indelicato_JPhysB2019,mohr:QEDrev,Sapirstein_PhysRevA2002,Sapirstein_PhysRevA.91.062508,Korobov_PhysRevLett2016,Korobov_PhysRevLett2017,Yerokhin_PhysLettA1995}. In recent years there has been an increased interest in the use of effective QED potentials\cite{shepler2005ab,Thierfelder2010PRA,Artemyev_chap2016,pavsteka2017relativistic}. An early example of such a potential is the Uehling potential for vacuum polarization\cite{Uehling1935}. The electron self-energy is more complicated to represent by a potential due to its delocal nature, but a number of such potentials are now available\cite{Pyykko2003a,Flambaum2005,Shabaev_PhysRevA.88.012513}. QED-effects have also been included in the fitting of relativistic effective core potentials\cite{hangele2012accurate,hangele2013relativistic}. We have implemented effective QED potentials in the DIRAC code for general relativistic molecular calculations\cite{saue2020dirac}. A full account of our implementation will be given elsewhere\cite{Sunaga2021qed}; very recently, a similar implementation has been reported by Skripnikov\cite{Skripnikov_JCP2021}. QED-effects on parity violation have previously been studied in  atoms and then with focus on transition amplitudes\cite{Blundell_PRD1992,johnson2001vacuum,milstein2002radiative,Shabaev_PRA2005,kuchiev2003radiative,Roberts_PRA2013a,Roberts_PRA2013b,Sahoo_MP2017,Yu_Khetselius_2017}. We believe that the present work is the first to study radiative corrections to molecular PV energies.

Our paper is organized as follows: In Section \ref{seq:effQED} we explain how effective QED potentials are included in our calculations. Section \ref{sec:PV} gives a brief introduction to
parity violation in chiral molecules. Computational details are given in in Section \ref{sec:comp}. In Section \ref{sec:results} we present and discuss our results, before concluding in Section \ref{sec:conc}. We employ SI-based atomic units throughout this paper.
\section{Theory}
\subsection{Relativistic Hamiltonian with effective QED potentials}\label{seq:effQED}

Our calculations are based on the Dirac--Coulomb Hamiltonian
\begin{equation}
{\hat H}_{\rm{DC}} = V_{nn} + \sum\limits_i {\hat h}(i) + \frac{1}{2}\sum\limits_{i \ne j}\frac{1}{r_{ij}}
\end{equation}
Here, the first and third terms represent the nucleus-nucleus and electron-electron repulsion, respectively. The one-electron part is given by the Dirac Hamiltonian
\begin{equation}
{\hat h} =   {\beta}m_{\rm{e}}c^2 + c\left(\boldsymbol{\alpha}\cdot{\bf p}\right) + V_{\rm{en}};\quad V_{\rm{en}}=-e\phi_n
\end{equation}
in the scalar potential $\phi_n$ of fixed nuclei and will be extended by effective QED-potentials $V_{\rm{effQED}}$
\begin{equation}
{\hat h} \rightarrow {\hat h} + V_{\rm{effQED}};\quad V_{{\rm{effQED}}} = \sum\limits_A\left(V^{\rm{VP}}_A + V^{\rm{SE}}_A\right)
\end{equation}
where $V^{\rm{VP}}_A$ and $V^{\rm{SE}}_A$ refer to the contributions from the vacuum polarization and the electron self-energy, respectively, associated with nucleus $A$.
We employ the Uehling potential \cite{Uehling1935} for the vacuum polarization term $V^{\rm{VP}}$.  The Uehling potential for a spherically symmetric extended nucleus can be expressed as follows \cite{Fullerton1976}
\begin{equation}
\label{eqn:Uehling_exact}
V^{{\rm{VP}}}_A(r) =  -\frac{2\alpha\lambdabar_e}{3}\frac{Z_Ae^2}{r}\int^{\infty}_0dr'r'\rho_A\left(r'\right)\left[K_0\left(\frac{2}{\lambdabar_e}\left|r-r'\right|\right)\right.  \left.-K_0\left(\frac{2}{\lambdabar_e}\left|r+r'\right|\right)\right]. 
\end{equation}
where
\begin{equation}
K_0(x)=\int^{\infty}_1 dt\exp{[-xt]}\left(\frac{1}{t^3}+\frac{1}{2t^5}\right)\left(t^2-1\right)^{1/2},
\end{equation}
$\alpha$ the fine structure constant, ${\lambdabar}_e$ the reduced Compton wavelength and $\rho_A$ is the normalized charge density of nucleus $A$.

For the self-energy term $V_{{\rm{SE}}}$, we have employed two kinds of effective potentials. The first one is the Gaussian model potential proposed by Pyykk\"{o} and Zhao (PZ) \cite{Pyykko2003a},
\begin{equation}
{V^{{\rm{SE}}}_A} \rightarrow {V^{\rm{PZ}}_A}\left( r \right) = B(Z_A)\exp \left( { - \beta (Z_A){r^2}} \right),
\end{equation}
where the parameters $B$ and $\beta$ are second-order polynomials in nuclear charge with coefficients obtained by fitting to SE data ($2s_{1/2}$ SE energy shift of hydrogen-like ions\cite{Indelicato_PhysRevA.58.165,Beier_PRA1998} and relative SE shift of hyperfine transition energies of lithium-like ions\cite{Boucard_EPJD2000}) in the range $29\le Z \le 83$.
%

The second effective SE potential was proposed by Flambaum and Ginges (FG) \cite{Flambaum2005}. It is formulated in the language of form factors expressing the effective interaction of an electron with the electromagnetic field, radiative corrections included\cite{Schweber_RQFT,Berestetskii:QED1982}. The potential is written as a sum of three contributions
\begin{equation}
{V^{{\rm{SE}}}_A} \rightarrow {V^{\rm{FG}}_A}\left( r \right)= V^\textrm{MAG}_A + V^\textrm{HF}_A + V^\textrm{LF}_A, 
\end{equation}
where the first term is associated with the magnetic form factor
\begin{equation}\label{eqn:MAG}
V^\textrm{MAG}_A(\mathbf{r})= \frac{\alpha\lambdabar_e}{4\pi} i \beta\boldsymbol{\alpha} \cdot \nabla \left[\phi_{n;A}(r) \left(\int^{\infty}_1dt \frac{1}{t^2 \sqrt{t^2-1}} e^{-\frac{2tr}{\lambdabar_e}}-1\right) \right]
\end{equation}
and contains the gradient of the scalar potential $\phi_{n;A}$ associated with nucleus $A$. The contribution from the electric form factor has been split into a high-frequency (HF) part
\begin{eqnarray}
\label{eqn:HF}
V^\textrm{HF}_A(r)= -A\left(Z_A,r\right)\frac{\alpha}{\pi} \phi_{n;A}(r) \int^{\infty}_1dt \frac{1}{\sqrt{t^2-1}}\left[ \left(1-\frac{1}{2t^2}\right)\right. \nonumber \\
\left.\times\left[\textrm{ln}\left(t^2-1\right)+4\textrm{ln}\left(\frac{1}{Z_A\alpha}+0.5\right)\right]-\frac{3}{2}+\frac{1}{t^2}\right]
e^{-\frac{2tr}{\lambdabar_e}},
\end{eqnarray}
obtained from theory after introducing a fixed low-frequency cutoff, and a low-frequency (LF) part
\begin{equation}\label{eqn:LF}
V^\textrm{LF}_A(r) = -B\left(Z_A\right) Z^4_A \alpha^5 m_e c^2 e^{-Z_Ar}.
\end{equation}
for which an approximate form has been chosen. Parameters $A\left(Z,r\right)$ and $B\left(Z\right)$ have been fitted to SE shifts of hydrogen-like ions \cite{Mohr_PhysRevA.45.2727,Mohr_PhysRevA.46.4421}.

These effective QED-potentials have been implemented by grafting routines from the numerical atomic code GRASP\cite{dyall1989grasp} onto the DFT-grid of the DIRAC code\cite{saue2020dirac,Saue_JCC2002,Sunaga2021qed}. Since the QED effects are principally generated in the vicinity of nuclei, on the order of reduced Compton wavelengths\cite{Artemyev_chap2016}, the numerical integration is only carried out on the nucleus associated with the QED-potential and with a radial cutoff. A problematic term in this respect is the low-frequency part of the Flambaum--Ginges potential since it by construction has the range of a hydrogen-like $1s$ orbital and for light elements may therefore overlap with other atoms in a molecule. Therefore, by default in the current implementation, QED-potentials are only activated for centers with $Z>18$.

The potentials can also be used at the eXact 2-Component relativistic (X2C) level\cite{ilias:jcp2007}. In the present work we employ the eXact 2-Component (X2C) molecular-mean field (mmf) Hamiltonian\cite{Sikkema_Visscher_Saue_Ilias_2009}. In this approach an initial HF calculation is carried out at the 4-component relativistic level. The converged Fock matrix is then transformed to 2-component form and employed in the ensuing correlated calculations.

\subsection{Parity-violating energy shift}\label{sec:PV}
Theoretical studies of parity violation in atoms and molecules are based on an effective Hamiltonian describing the weak interaction between electrons and nucleons mediated by the exchange of virtual $Z^0$ bosons. A detailed derivation and discussion of this Hamiltonian, starting at the level of quarks, is for instance found in Ref.~\cite{Greiner_weak} (the reader may also consult Refs.~\citenum{Berger:Parity-violation_effects_in_molecules,bast2011analysis,Bakasov_JCP_1998,halzen:quark,Hung_Sakurai_ARNPS1981,Commins_Bucksbaum_ARNPS1980}). The $Z^0$ boson is neutral, like its electroweak partner, the photon. On the other hand, it has mass --- 91.1876(21) GeV/c \cite{nakamura2010particle}, which is close to that of $^{98}\textrm{Mo}$, the most abundant isotope of molybdenum \cite{bast2011analysis}. This confers a short-range nature to the interaction. The effective Hamiltonian accordingly has the form of a contact interaction between two fermion 4-currents,
\begin{equation}
  \hat{h}_{\rm{eff}}\sim\int j^{\mu;e}\left(\boldsymbol{r},t\right)\left(\sum_{i}^{Z}j_{\mu;i}^{p}\left(\boldsymbol{r},t\right)+\sum_{i}^{N}j_{\mu;i}^{n}\left(\boldsymbol{r},t\right)\right)d^{3}\boldsymbol{r}
\end{equation}
in line with the form suggested by Fermi in 1934\cite{Fermi_ZfP1934}. The fermion currents of electromagnetic interaction have the form $\bar{\psi}\gamma^{\mu}\psi=\psi^{\dagger}(I_4,\boldsymbol{\alpha})\psi$, which is termed a vector current since the space-like part of the current transforms as a vector under space inversion. The fermion currents associated with the weak interaction are of mixed vector and axial vector (V-A) nature\cite{Sudarshan_PR1958,Feynman_PR1958,Sakurai_NC1958}, the latter on the form $\psi^{\dagger}(\gamma^5,\boldsymbol{\Sigma})\psi$, where appears 
\begin{equation}
\gamma^{5}=\left[\begin{array}{cc}
0_{2} & I_{2}\\
I_{2} & 0_{2}
\end{array}\right].
\label{eqn:gamma5}
\end{equation}
For the nucleon currents a non-relativistic approximation is used, setting the small components of the 4-spinors to zero, which in practice means that terms involving the odd operators $\gamma^5$ and $\boldsymbol{\alpha}$ vanish whereas even operators are reduced to their 2-component counterparts, that is,  $I_4,\boldsymbol{\Sigma}\rightarrow I_2,\boldsymbol{\sigma}$.

Parity-violation arises from the $V_e-A_{p,n}$ and $A_e-V_{p,n}$ cross-terms. The former coupling leads to a nuclear spin-dependent effective Hamiltonian, which has so far been ignored in the calculation of parity-violation energies, but has been employed in studies of parity-violation in the NMR spectra of chiral molecules\cite{Barra_PLA1986,Barra_EPL1988,Laubender_CPC2003,Soncini_PhysRevA.68.033402,Bast2006JCP,Nahrwold_JCP2009}. Here we limit attention to the latter coupling, which leads to a nuclear spin-independent effective Hamiltonian on the form
\begin{equation}
\hat{h}^{\rm{PV}}_{\rm{eff}} = \frac{{{G_{\rm{F}}}}}{{{\rm{2}}\sqrt 2 }}\sum\limits_A {Q_{W,A}}\sum\limits_i {\gamma_i^5\rho_A\left({\mathbf{r}_i}\right)} .
\label{eqn:pv_op}\end{equation}
where 
\begin{equation}
Q_{W,A} =  - N_A + Z_A\left( {1 - 4{{\sin }^2}{\theta_W }} \right),
\end{equation}
is the weak charge\cite{Bouchiat_JPF1974} of nucleus $A$ given in terms of the numbers $N_A$ and $Z_A$ of neutrons and protons, respectively, as well as the weak mixing angle $\theta_\textrm{W}$ \cite{glashow1961partial,weinberg1967model}. The current recommended value\footnote{The value depends on the renormalization scheme employed. The PDG employs the modified minimal subtraction scheme, whereas the conceptually simpler, but as such less accurate\cite{CODATA2002} CODATA value $\sin^2\theta_\textrm{W}=1-(m_W/m_Z)^2=0.22290(30)$ \cite{CODATA2018}, given explicitly in terms of the masses of the $Z^0$ and $W^{\pm}$ bosons is obtained from the on-shell scheme.} from the Particle Data Group (PDG)\cite{PDG2020} is $\sin^2\theta_\textrm{W}=0.231 22(4)$.

The presence of the Fermi coupling constant $G_\textrm{F}=2.22255\times 10^{-14}E_ha_0^3$ in the effective Hamiltonian \eqref{eqn:pv_op} shows that the interaction is truly weak. The operator can therefore not be added to the electronic Hamiltonian in an energy calculation, since its contribution would be masked by numerical noise, rather it will be obtained perturbatively as an expectation value. The expectation value with respect to the electronic wavefunction $\Psi$ can in turn be expressed as a sum over atomic contributions,
\begin{equation}
E^{\rm{PV}} = \sum\limits_A  E^{\rm{PV}}_A =  \frac{G_F}{2\sqrt 2}\sum\limits_A Q^A_{W} M_A^{\rm{PV}};
\quad 
M_A^{\rm{PV}} = \sum\limits_i\left<\Psi \left| \gamma_i^5\rho_A\left({\mathbf{r}_i}\right) \right|\Psi\right>.
\end{equation}
The reduced PV matrix element $M_A^{\rm{PV}}$ is of highly local nature, not only because of the presence of  the normalized nuclear charge density $\rho_A$ of each center $A$, but also because the $\gamma_5$ matrix \eqref{eqn:gamma5} couples large and small components, the latter being of highly local atomic nature for positive-energy orbitals.

The parity-violating energy $E^{\rm{PV}}$ is strictly zero in the absence of spin-orbit coupling and its meaningful calculation in a non-relativistic framework hence requires double perturbation theory\cite{Gajzago_AIPConf1974,Rein_PLA1979,hegstrom1980calculation,mason1984parity,lazzeretti1997calculation,Bakasov_JCP_1998,berger2000multiconfiguration}. The non-relativistic form of the effective Hamiltonian \eqref{eqn:pv_op} and spin-orbit interaction then contributes factors $Z_A^3$ and $Z_B^2$, respectively, giving an overall $Z^5$  scaling. Single-center contributions ($A=B$) vanish in a non-relativistic framework\cite {hegstrom1980calculation}, but not necessarily in a relativistic setting\cite{bast2011analysis}.

In a previous study, we analyzed the parity-violating energy of the \ce{H2X2}-series of molecules at the Hartree--Fock(HF) and Kohn--Sham(KS) levels\cite{bast2011analysis}. The $E^{\rm{PV}}$ expectation value was decomposed in terms of pre-calculated atomic orbitals. It was
found that a contribution $E^{\rm{PV}}_A$ is completely dominated by contributions from the same center $A$. Inserting the form of relativistic
atomic orbitals
\begin{equation}
  \psi(\mathbf{r})=\left[
  \begin{array}{r}R^L(r)\chi_{\kappa,m_j}(\theta,\phi)\\iR^S(r)\chi_{-\kappa,m_j}(\theta,\phi)\end{array}
  \right],
\end{equation}
the expectation value $E^{\rm{PV}}_A$ can be expressed in terms of products of radial and angular contributions. 
From the angular part of  it was found that only pairs of atomic orbitals with \textit{opposite} values of $\kappa$ and the \textit{same} value of $m_j$ can contribute. From the small $r$ behaviour of the radial functions it was found that for point charge nuclei only $(s_{1/2},p_{1/2})$ - pairs of atomic orbitals can contribute, whereas for extended nuclei contributions with $|\kappa|>1$ are allowed. Finally, for the \ce{H2X2}-series of molecules we found that the reduced PV matrix element $M^X_{\rm{PV}}$ can be remarkably well approximated by the mixing of valence $ns_{1/2}$ and $np_{1/2}$ orbitals on the same center $X$, that is
\begin{equation}
  M^X_{\rm{PV}}\approx M^X_{\rm{PV};\it{nn}}\times 2\rm{Re}\left[\sum_{\it{i}} \it{c}(\it{ns}_{\rm{1/2}}^X)^\ast_i \it{c}(\it{np}_{\rm{1/2}}^X)_i\right],\label{eq:MX_PV}
\end{equation}
where
\begin{equation}
M^X_{\rm{PV};\it{mn}}=\langle ms_{1/2}^X|\gamma_5\rho^X|np_{1/2}^X\rangle =  \langle R_{ms_{1/2}}^{L;X}|\rho^X|R_{np_{1/2}}^{S;X}\rangle_r + \langle R_{ms_{1/2}}^{S;X}|\rho^X|R_{np_{1/2}}^{L;X}\rangle_r .\label{eq:PVint}
\end{equation}
In \eqref{eq:MX_PV} $c(\psi)_i$ is the coefficient of atomic orbital $\psi$ in the expansion of molecular orbital $i$. The quantity $2\rm{Re}\left[\sum_{\it{i}} \it{c}(\it{ns}_{\rm{1/2}}^X)^\ast_i \it{c}(\it{np}_{\rm{1/2}}^X)_i\right]$ thereby determines the mixing of atomic orbitals $ns_{1/2}$ and $np_{1/2}$ \textit{contributing} to the PV energy, but it should be noted that this particular form refers to a specific phase convention for atomic orbitals. Further details are found in Ref.~\citenum{bast2011analysis}.

\section{Computational details}\label{sec:comp}
Unless otherwise stated all calculations have been carried out using the DIRAC program \cite{saue2020dirac} (revisions 9e12f6d89, 13d3f86a3, 9533dd59e, 09012db84, and ccb40d212).
We employed a Gaussian model for the nuclear charge distribution in all calculations\cite{visscher1997dirac}.
Molecular geometries were taken from Table I of Ref.~\citenum{laerdahl1999fully} and were optimized at the MP2 level using quasirelativistic pseudopotentials for the heavy elements.
In all calculations the dihedral angle was fixed at 45$^{\circ}$ of the P enantiomer. For the weak mixing angle, we have used the value $\sin^2\theta_\textrm{W}=0.2319$ in line with previous works\cite{Bakasov_JCP_1998,laerdahl1999fully} and corresponding to the PDG 1994 recommended value\cite{PDG_1994}.

Various flavours of Dyall basis sets\cite{dyall2002relativistic, dyall2006relativistic, dyall2016relativistic} have been employed for all elements of the target molecules. These basis sets have been extended by tight exponents for better description of the PV operator.
These optimizations were carried out using the simplex method\cite{pressnumerical}, minimizing the target function
\begin{equation}\label{eq:PVerr}
\Delta M_{\rm{PV}}\left( \%  \right) = 100 \times \frac{1}{N}\sum\limits_{nm} {{\rm{abs}}\left( \frac{M_{\rm{PV,}\it{nm} }^{{\rm{num}}} -  M_{\rm{PV,}\it{nm} }^{\rm{bas}}}{M_{\rm{PV,}\it{nm} }^{\rm{num}}} \right)} ,
\end{equation}
where $M_{\rm{PV,}nm }^{{\rm{num}}}$ and $M_{\rm{PV,}nm }^{\rm{bas}}$ refer to reduced PV matrix elements, \eqref{eq:PVint}, calculated numerically or in the current basis set, respectively.  
For this, we employed a development version of the atomic GRASP code\cite{dyall1989grasp} allowing the use of Gaussian basis sets, in this case the Dyall.$n$zp ($n=2,3,4$) basis sets for SCF calculations. 

For the MP2 and CCSD calculations a computational protocol was developed based on a calibration study of the parity-violating energy $E_{PV}$ of the \ce{H2Se2} molecule, 
described in the Supplemental Material. All calculations employed the molecular-mean-field X2C Hamiltonian ($^2DC_M$) \cite{Sikkema_Visscher_Saue_Ilias_2009} based on the  Dirac--Coulomb  Hamiltonian. For technical reasons, MP2 and CCSD analytical expectation values were calculated using the RELCCSD\cite{Visscher_Lee_Dyall_1996,shee2016analytic} and ExaCorr\cite{pototschnig2021implementation} modules, respectively. The expectation
values were calculated using the Lagrangian-based analytical energy derivative technique described in Ref.~\citenum{shee2016analytic} and references therein.
For the CCSD calculations, we used the dyall.aae3z basis sets \cite{dyall2002relativistic, dyall2006relativistic, dyall2016relativistic}, extended with tight functions, as described above. We correlated $\textrm{(}n-1\textrm{)}s\textrm{(}n-1\textrm{)}p\textrm{(}n-1\textrm{)}dnsnp$ electrons, and truncated the virtual space at 100 $E_h$.
For MP2 the calibration study indicated that only valence orbitals need to be correlated, so we used the the dyall.acv3z basis.

For comparison we carried out DFT calculations using the extended dyall.aae3z basis sets and functionals from different rungs of the Jacob's ladder of density functional approximations\cite{perdew:jacob}: LDA (SVWN5)\cite{Slater1951LDA, Vosko1980LDA}, BLYP\cite{Becke1988BLYP, Lee1988BLYP, Miehlich1989BLYP}, and B3LYP\cite{Stephens_Devlin_Chabalowski_Frisch_1994,Becke1993B3LYP,Hertwig1997B3LYP}, which is Becke's series, and PBE\cite{Perdew1986PBE, Perdew1996PBE} and PBE0\cite{Adamo1999PBE0, Ernzerhof1999PBE0}, which is Perdew's series. We also used the long-range correlated functional CAMB3LYP\cite{Yanai2004}.

\section{Results and discussion.}\label{sec:results}
For convenience, we have collected our HF, MP2 and CC results in Table 6 of the Supplemental Material together with previous results\cite{Thyssen_PRL2000,van2005first,shee2016analytic} calculated at the same geometries\cite{laerdahl1999fully}. Our outputs are available via the zenodo repository\cite{paper:dataset}. 

\subsection{Optimization of tight exponents}\label{Sec:simplex}
In Table \ref{Tab:error_M_PV} we show the average relative error $\Delta M_{\rm{PV}}$  \eqref{eq:PVerr} of reduced PV matrix elements for various basis sets with respect to the reference numerical calculation. For Se, we see a reduction from 19.8\% to 3.2\% when going from dyall.2zp to dyall.4zp unmodified basis sets. At the dyall.3zp level the relative error is reduced for the heavier elements since these basis sets already contain quite tight exponents. 
Adding tight exponents, however, the error can be brought below 1\% at all basis set levels.
Specifically, we added $2p$ for O, $1s3p$ for S and $1p$ for Te at the dyall.3zp level. For the Se atom, we added $1s3p$ for dyall.2zp, $2p$ for dyall.3zp, and $1p$ for dyall.4zp, respectively. The values of the optimized exponents are given in the Supplemental Material. For Po no exponents were added since the relative error was already well below 1\%. 

\begin{table}[]
\tbl{Average relative error  $\Delta M_{\rm{PV}}$(\%) in reduced PV matrix elements in different basis sets (with and without added tight exponents) with respect to the reference numerical values.}
{\begin{tabular}{lcccccccc}
\hline
 & O & S & \multicolumn{3}{c}{Se} & Te & Po\tabularnewline
basis & dyall.3zp & dyall.3zp & dyall.2zp & dyall.3zp & dyall.4zp & dyall.3zp & dyall.3zp\tabularnewline \hline
without & 11.4 & 11.4 & 19.8 & 8.5 & 3.2 & 3.4 & 0.1\tabularnewline
with & 0.9 & 0.5 & 0.9 & 0.6 & 0.2 & 0.1 & \tabularnewline
\hline
\end{tabular}}
\label{Tab:error_M_PV}
\end{table}


\subsection{Comparison of methods}
Parity-violating energies of the \ce{H2X2} molecules, calculated at the 45$^{\circ}$ dihedral angle of the P enantiomer using different methods, are  collected in Table \ref{Tab:DFT}. Using the CCSD results as reference, we find that the mean relative error (MRE) of the HF method, corresponding to the relative correlation effect, is -2.8\%, so rather modest, in line with previous studies\cite{Thyssen_PRL2000,shee2016analytic}. On the other hand, the standard deviation $\Delta_{\rm{std}}$ approaches ten percent, showing quite a spread. MP2 has a slightly smaller MRE of opposite sign (2.0 \%) and similar spread. No density functional outperforms HF; their average MRE and $\Delta_{\rm{std}}$ are -11.1 \% and 7.0\%, respectively. The best performing functional is PBE0 (-5.7$\pm$ 5.0 \%), the worst CAMB3LYP (-17.2$\pm$4.3 \%). These observations are as such disappointing since, as already mentioned, the computational protocol developed for the scan of candidate molecules for the Paris experiment is based on DFT. On the other hand, the protocol calculates a vibrational shift, corresponding to the PV energy difference of two vibrational levels, rather than an absolute electronic PV energy as here. Further study is therefore required to judge the performance of the computational protocol. 

\begin{table}[h]
\tbl{Summary of the values of $E_{\rm{PV}} (E_h)$, calculated at the 45$^{\circ}$ dihedral angle of the P enatniomer of \ce{H2X2} molecules at various levels of theory. "err" is the percentage error with respect to CCSD. }
{\begin{tabular}{lrrrrrrrrrr}
\hline
 & \multicolumn{2}{c}{\ce{H2O2}} & \multicolumn{2}{c}{\ce{H2S2}} & \multicolumn{2}{c}{\ce{H2Se2}} & \multicolumn{2}{c}{\ce{H2Te2}} & \multicolumn{2}{c}{\ce{H2Po2}}\tabularnewline
 Method & $E_\textrm{PV}$ &  err (\%) & $E_\textrm{PV}$ & err (\%) & $E_\textrm{PV}$ & err (\%) & $E_\textrm{PV}$ & err (\%) & $E_\textrm{PV}$ & err (\%) \tabularnewline \hline
CCSD     & -5.917E-19 &       & -2.166E-17 &         & -2.463E-15 &       & -3.536E-14 &       & -1.337E-12 &      \tabularnewline
MP2      &  -6.525E-19 &  -10.3 & -2.067E-17 &  4.6 & -2.191E-15 & 11.1  & -3.236E-14 &   8.5 &  -1.390E-12&  -3.9\tabularnewline
HF       & -6.808E-19 & -15.0 &  -2.160E-17 & 0.3 & -2.301E-15 &  6.6  & -3.403E-14 &   3.8 & -1.467E-12 &   -9.7\tabularnewline
LDA      & -6.581E-19 & -11.2 & -2.659E-17 &  -22.7& -2.910E-15 & -18.2 & -4.164E-14 & -17.7 & -1.308E-12 &    2.2\tabularnewline
BLYP     & -6.303E-19 & -6.5  & -2.663E-17 & -22.9& -2.864E-15 & -16.3 & -4.095E-14 & -15.8 & -1.324E-12 &    1.0\tabularnewline
B3LYP    & -6.515E-19 & -10.1 & -2.576E-17 &  -18.9 & -2.796E-15 & -13.5 & -4.032E-14 & -14.0 & -1.372E-12 &   -2.6\tabularnewline
CAMB3LYP & -6.627E-19 & -12.0 & -2.621E-17 & -21.0& -2.941E-15 & -19.4 & -4.263E-14 & -20.5 & -1.515E-12 &  -13.3\tabularnewline
PBE      & -6.269E-19 & -5.9  & -2.498E-17 & -15.3 & -2.653E-15 & -7.7  & -3.814E-14 &  -7.8 & -1.261E-12 &    5.7\tabularnewline
PBE0     & -6.487E-19 & -9.6  & -2.404E-17 & -11.0 & -2.570E-15 & -4.4  & -3.729E-14 &  -5.4 & -1.313E-12 &    1.8\tabularnewline
\hline
\end{tabular}}
\label{Tab:DFT}
\end{table}

\subsection{Radiative corrections to PV energies}

In Table \ref{Tab:QED} we show QED-effects on parity-violating energies for the heavier members of the \ce{H2X2}-series, calculated at the CCSD level. We used the Uehling term (U) for the inclusion of vacuum polarization (VP) and the Pyykk{\"o}--Zhao (PZ) or Flambaum--Ginges (FG) potentials for electron self-energy (SE). Fitting our data to a simple power law $\sim Z_X^\gamma$ with respect to the nuclear charge of the heavy atom $X$, we find that the uncorrected PV energy has exponent 7.0, whereas the radiative corrections show stronger scaling, with $\gamma$ equal to 8.1 and 7.7, for $\Delta$(U+PZ), and $\Delta$(U+FG), respectively. In passing we note that this trend may not hold for superheavy systems, since SE and VP terms tend to have opposite sign, as known for valence atomic properties (see for instance Refs.~\citenum{pyykko1998estimated,Thierfelder2010PRA}). 

From Table \ref{Tab:QED} we see that the radiative correction may reach a few percent of the PV energy, but only if we look at the results obtained with the FG potential. The discrepancy between the results obtained with the PZ and FG potentials could signal that one or both are being used beyond their domain of validity, since they have been principally designed to reproduce QED energy shifts.

\begin{table}[]
  \tbl{QED effects on the parity-violating energy $E_\textrm{PV}$ ($E_h$), calculated at the 45$^{\circ}$ dihedral angle of the P enantiomer of $\textrm{H}_2\textrm{X}_2$ molecules.
The radiative corrections have been included using the Uehling potential (U) for vacuum polarization and either the Pyykk\"{o}--Zhao (PZ) or Flambaum--Ginges (FG) effective potentials for electron self-energy. Absolute and relative differences are also shown.}{\begin{tabular}{lccccccc}
\hline
 & non-QED & U+PZ & U+FG & $\Delta$(U+PZ)($E_h$) & $\Delta$(U+FG)($E_h$) & $\Delta$(U+PZ)(\%) & $\Delta$(U+FG)(\%) \tabularnewline  \hline
\ce{H2Se2} & -2.463E-15 & -2.458E-15 & -2.434E-15 & 4.6E-18 & 2.9E-17 & -0.19 & -1.18\tabularnewline
\ce{H2Te2} & -3.536E-14 & -3.525E-14 & -3.471E-14 & 1.2E-16 & 6.6E-16 & -0.33 & -1.89\tabularnewline
\ce{H2Po2} & -1.337E-12 & -1.331E-12 & -1.306E-12 & 6.8E-15 & 3.1E-14 & -0.51 & -2.38\tabularnewline
\hline
\end{tabular}}
\label{Tab:QED}
\end{table}



\section{Conclusions}\label{sec:conc}
In this article we have given improved values for the parity violating energy of \ce{H2X2} molecules (X = O, S, Se, Te, Po). Improvements include: i) Property-optimized basis sets, which go beyond conventional energy-optimized basis sets. ii) We have used a new relativistic coupled cluster code, ExaCorr, geared towards massively parallel calculations\cite{pototschnig2021implementation}, allowing extensive calibration studies in order to elaborate a CCSD computational protocol for PV energies and to include \ce{H2Po2} among the target molecules iii) Inclusion of QED-effects through effective potentials.

Using our CCSD numbers as reference, we find the performance of DFT somewhat disappointing. This is possibly worrisome, since DFT is the basis of the computational protocol elaborated for the scan of candidate molecules for the high-resolution laser spectroscopy experiment which is under development in Paris and which aims to provide the first observation of parity violation in molecules as a shift in the vibrational spectra of enantiomers of chiral molecules. We shall pursue our investigation of the reliability of this protocol by calculating at the coupled cluster level the vibrational shift, involving energy differences, of candidate molecules. On the positive side, it has been noted that HF, here better performing, tends to predict larger PV vibrational shifts than DFT \cite{DeMontigny_PCCP2010}.

A major novelty of this work is the first calculation of molecular PV energies $E_{\rm{PV}}$ including QED effects. For the heaviest molecules in the \ce{H2X2}-series the radiative corrections are on the order of a few percent. However, we do note that this depends on what effective self-energy potential is used. This may indicate limitations to their applicability, an issue that encourages us to develop more rigid approaches to incorporate QED effects. 

Further improvements would include going beyond the zeroth-order term, that is, the instantaneous Coulomb term, of the fully relativistic electron-electron interaction\cite{saue2011relativistic}.
The contribution from the Breit term, or more precisely spin-other-orbit interaction, has been estimated by Berger within the Breit--Pauli framework\cite{Berger_JCP2008}. The Breit effects for \ce{H2X2} (X = O, S, Se, Te, Po) are 12\%, 4.7\%, 1.9\%, 1.2\%, 0.7\%, respectively and hence monotonically decreasing along the series, contrary to the radiative corrections.

Another improvement would be to increase the excitation level beyond that of CCSD. An indication of the role of higher-order correlation effects comes from the finite-field study of Thyssen \textit{et al}\cite{Thyssen_PRL2000}: The relative difference between CCSD(T) and CCSD was found to be $-1.0 \%$ and $-2.5 \%$ for \ce{H2O2} and \ce{H2S2}, respectively. This suggests that these effects may be at the percentage level and increase down the series. This clearly calls for further study.

\section*{Acknowledgement(s)}
This paper is dedicated to John F. Stanton, a driving force of coupled cluster theory for 35 years !
We would like to thank David Sanchez(Toulouse) and Johann Pototschnig(Vienna) for a lot of technical support.
This work is part of the project molQED funded by the Agence Nationale de la Recherche (ANR, France).
This research used resources of the Oak Ridge Leadership Computing Facility, which is a DOE Office of Science User Facility supported under Contract DE-AC05-00OR22725. We are thankful to CALMIP (Calcul en Midi-Pyr\'{e}n\'{e}es) and LCPQ (Laboratoire de Chimie et Physique Quantiques) clusters for the additional computation time. We are also thankful to the supercomputer of ACCMS (Kyoto University) and Research Institute for Information Technology, Kyushu University (General Projects) for test calculations. 
A.S. acknowledges financial support from the Japan Society for the Promotion of Science (JSPS) KAKENHI Grant No. 17J02767 and 20K22553, and JSPS Overseas Challenge Program for Young Researchers, Grant No. 201880193.

%





\bibliography{article}
\bibliographystyle{tfo}
\end{document}